\documentclass[a4paper,fleqn]{cas-dc}

\usepackage[numbers]{natbib}

\usepackage{xurl} 
\usepackage{hyperref}
\usepackage{color}
\usepackage{subfigure}
\usepackage{floatrow}
\usepackage{graphicx,epstopdf}
\usepackage{ragged2e}
\usepackage{mciteplus}
\usepackage{chngcntr}
\usepackage[english]{babel}
\usepackage{changepage}
\DeclareMathOperator\erf{erf}
\usepackage{caption} 
\usepackage{threeparttable}
\def\tsc#1{\csdef{#1}{\textsc{\lowercase{#1}}\xspace}}
\tsc{WGM}
\tsc{QE}
\tsc{EP}
\tsc{PMS}
\tsc{BEC}
\tsc{DE}

\begin{document}
\let\WriteBookmarks\relax
\def\floatpagepagefraction{1}
\def\textpagefraction{.001}
\shorttitle{Methods for forecasting the effect of exogenous risk on stock markets}
\shortauthors{K Arias Calluari et~al.}

\title [mode = title]{Methods for forecasting the effect of exogenous risks on stock markets}                      
\author[1]{Karina Arias-Calluari}
\fnmark[1]
\address[1]{School of Civil Engineering, The University of Sydney, Australia}

\author[1]{Fernando Alonso-Marroquin}

\author[2]{Morteza. N. Najafi}
\address[2]{Department of Physics, University of Mohaghegh Ardabili, Ardabil, Iran}

\author[3]{Michael Harr\'e}
\address[3]{Complex Systems Research Group, Faculty of Engineering, The University of Sydney, Australia}

\fntext[fn1]{ https://orcid.org/0000-0001-6013-5490 }
\nonumnote{karina.ariascalluari@uni.sydney.edu.au}

\begin{abstract}
Markets are subjected to both endogenous and exogenous risks that have caused disruptions to financial and economic markets around the globe, leading eventually to fast stock market declines. 
In the past, markets have recovered after any economic disruption. On this basis, we focus on the outbreak of COVID-19 as a case study of an exogenous risk and analyze its impact on the Standard and Poor's 500 (S\&P500) index. We assumed that the S\&P500 index reaches a minimum before rising again in the not-too-distant future. Here we present two cases to forecast the S\&P500 index.  The first case uses an estimation of expected deaths released on 02/04/2020 by the University of Washington. For the second case, it is assumed that the peak number of deaths will occur 2-months since the first confirmed case occurred in the USA. The decline and recovery in the index were estimated for the following three months after the initial point of the predicted trend. The forecast is a projection of a prediction with stochastic fluctuations described by $q$-gaussian diffusion process with three spatio-temporal regimes. Our forecast was made on the premise that any market response can be decomposed into an overall deterministic trend and a stochastic term. The prediction was based on the deterministic part and for this case study is approximated by the extrapolation of the S\&P500 data trend in the initial stages of the outbreak. The stochastic fluctuations have the same structure as the one derived from the past 24 years. A reasonable forecast was achieved with 85\% of accuracy.\\ 
\end{abstract}


\maketitle

\section{Introduction}
In the investigation of the stock market dynamics, there are two well-defined approaches: the ``descriptive" and ``structural" models \cite{anand2018structural}. Nowadays, these models attempt to capture the {\it endogenous} and {\it exogenous} systemic risks of market movements\cite{anand2018structural,patzelt2018universal,puertas2020stock,molina2020market}.  {\it Endogenous} risks have been modeled in a microeconomic fashion by considering the interactions between agents and how these change over time, causing non-linear disruptions even if system parameters change in a smooth fashion~\cite{wolpert2012hysteresis,harre2014strategic,chen2019spatial}. New ``descriptive models" present {\it endogenous} systemic risks as a dynamic factor affected by the flow of market orders and time scale \cite{patzelt2018universal}, excessive profits, and excessive losses \cite{denys2016universality} and large positive or negative variation in stock markets index \cite{valenti2018stabilizing,chowdhury2020predicting} due to market activity. The recent ``structural models" which are based on physical model systems such as a combination of oscillation within a basin of free energy or external force \cite{puertas2020stock,garcia2020forecast},  spin glasses \cite{kuyyamudi2019emergence}, and kinetic Ising model to model stock market network \cite{hoang2019data}  proposed a natural non-equilibrium system. In most cases, they introduce the {\it exogenous} systemic risks as a global risk component which evolves all the time affecting prices of traded assets. The effects of these systemic risks are likely causes of abrupt changes at the macro-level of market dynamics~\cite{onnela2003dynamic,harre2015entropy,deev2020connectedness,chen2020correlation}.
  
 In both systemic risks ({\it endogenous} and {\it exogenous}) the market response can be decomposed into two parts, a response function that models changes in the overall trend of the system, called the skeleton~\cite{hommes2001financial}, and the other is the stochastic term \cite{feng2020generalized}. In this work, we analyse the nonlinear properties of both of these terms. 
 
 The {\it endogenous} systemic risks are an inherent part of the nonlinear dynamics of a market and may have detectable precursor signals that act as warnings similar to those used in other nonlinear systems such as climate and ecology \cite{lenton2011early}. As an example, the 1987 market crash is likely an endogenous event \cite{sornette2006endogenous}, as it had a measurably different effect on market dynamics than the September 11 attacks~\cite{harre2009phase} and the 1995 Kobe earthquake \cite{sornette2006endogenous} which pose systemic {\it exogenous} risks to markets. The last two events cannot be endogenized into market prices by `rational' agents ahead of time because they are not foreseeable \cite{buccheri2013evolution}. 
 
 This paper presents the solution of a critical outstanding problem in finding how the market is coupled to the globe, and how exogenous and unpredictable global events produce deterministic trends in the market. In Section \ref{Sec:Model}, we decompose the price return into a deterministic component and a $q$-stationarity fluctuating term. In Section \ref{Sec:Covid19}, the solution was applied to the forecasting of S\&P500 index's response due to the outbreak of COVID-19. In Section \ref{Sec:Discussion}, we present a discussion on how this model provides an overview of the impacts of an exogenous market shock that can be used as a reference in developing predictive models with more accuracy.

\section{The Model} \label{Sec:Model}

\begin{figure}[!htbp]
	\centering
	\includegraphics[scale=0.4000,trim=0cm 0cm 3cm 1cm]{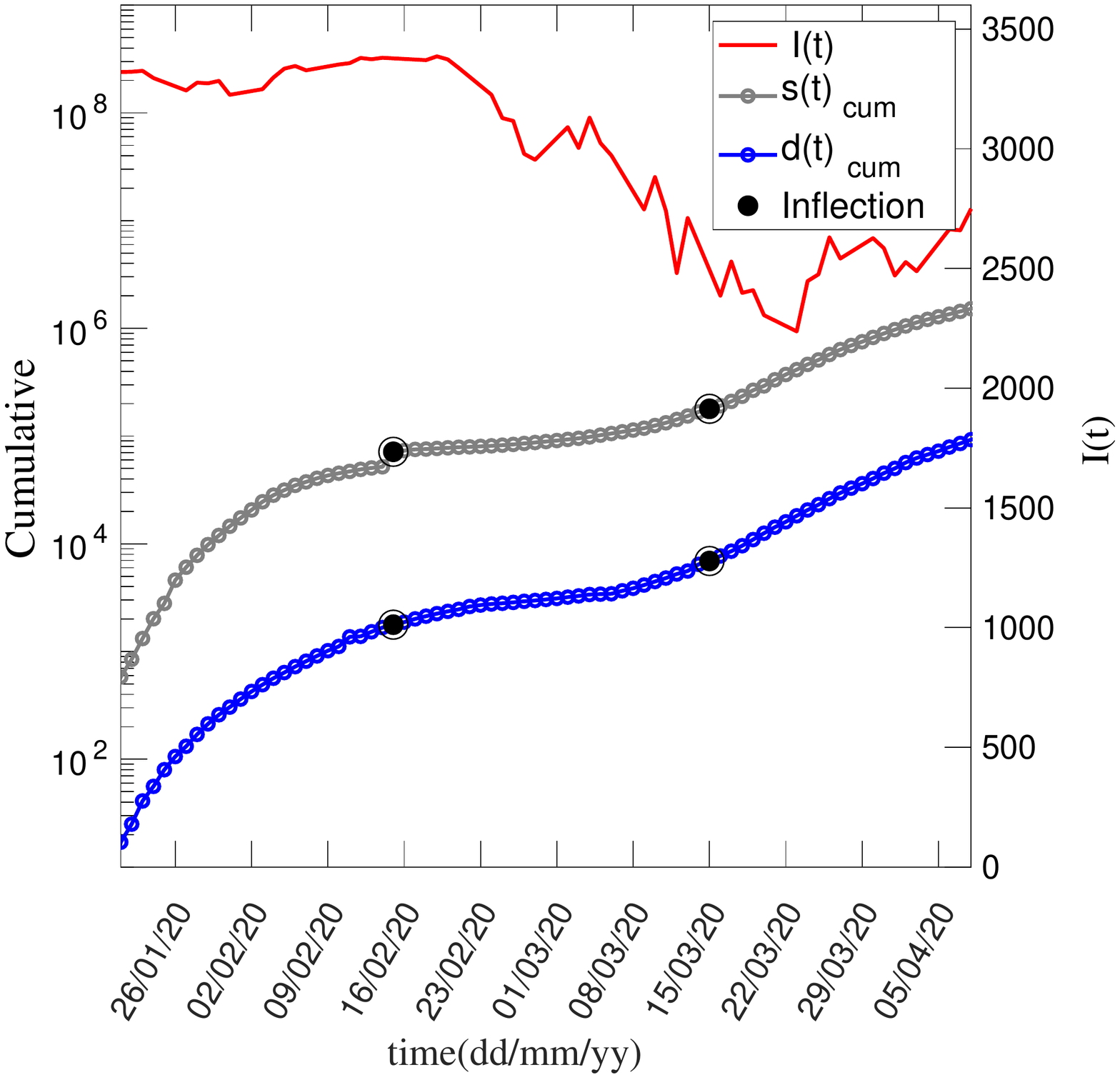}
	\caption{The S\&P500 decreased after the WHO (World Health Organization) reported an increment in confirmed cases $s(t)$ and number of deaths $d(t)$ due to COVID-19. The increments in the number of cases are represented with a black filled point, which are an abrupt change of slope of the first derivative of $s(t)_{cum}$ and $d(t)_{cum}$.  }
	\label{fig:Market_Lag}       
\end{figure}

\textcolor{black}{For this analysis, it is assumed that COVID-19 poses systematic \textit{exogenous} risks that affect stock markets behaviour. Therefore, our forecast models considered epidemiological research results to project the impact of COVID-19. Epidemiological researchers from around the world have produced an extensive array of analyses in order to model the spread, growth, peak, and ultimately the decline of the disease \cite{anderson2020will,chang2020modelling,neher2020potential,li2020retrospective,Ludin2020economic}}. For countries like China, Japan, Italy, and Iran, their epidemiological curve of COVID-19 progression displays a peak before the second month since the first cases were detected \cite{surveillances2020epidemiological,fanelli2020analysis,abdi2020coronavirus}. In countries like the UK, Australia and Germany, the governments have taken mitigation measures to slow the impact \cite{chang2020modelling,anderson2020will}.\textcolor{black}{The simultaneous reaction of governments, companies, consumers and media, have created a demand and supply shock, making COVID-19 a qualitatively different economic crisis than previous crises \cite{fernandes2020economic}}. This economic `wedging' of falling supply and demand is caused by rapidly escalating unemployment decreasing consumer demand where, in the US for example, 17 million Americans have applied for unemployment insurance in the first three weeks of the crisis \cite{Chaney2020}, and businesses are closing their doors reducing the supply of manufactured goods across the globe \cite{Donnan2020}. As a consequence financial markets around the world have fallen precipitously and market volatility is at near-all-time highs \cite{Watts2020}. For example the S\&P500 has registered the worst one-day fall in the last $24$ years and the third biggest percentage loss in its history.\\
In the uncertain environment of COVID-19 it is difficult to forecast the fluctuations of the S\&P500 index. We then need assumptions such as the mortality rates due to COVID-19 or the duration of the current shutdown of economic activity. Our central assumption is based on a predicted peak for COVID-19 deaths. Current results in Figure \ref{fig:Market_Lag} based on the daily World Health Organization reports \cite{WHS} show that the S\&P500 responds to the inflection points of the cumulative amount of deaths and confirmed cases with no lag. This fact supports our assumption that when a peak number of deaths occurs, the market reaches a stationary point.  

\textcolor{black}{We developed two cases to forecast the fluctuations of the S\&P 500. In the first case the peak number was located on the 23rd of May by the University of Washington \cite{UW}. The second case is where the peak number of deaths is considered 2-months since the first death occurred \cite{lai2020severe}. After this point we expect a recovery period as economic activity starts to return to normal.}
To construct the forecast, we assume that the stock market index can be decomposed into a deterministic trend and a stationary stochastic fluctuation. The statistics of the fluctuation have been obtained by analyzing the S\&P500 index during the 24 year period from January 1996 to March 2020 \citep{arias2019stationarity}. $I(t_0)$ is the initial point of the stock market index for some time point $t_0$, and the index $I(t)$ for $t>t_0$ is its time evolution. In these predictions we take $t_0=24/03/2020$ and $t_0=28/02/2020$ for the two different forecast. The price return at time $t$ is defined by: \\
\begin{equation}\label{eq:Price return}
X(t)=I(t_{0}+t)-I(t_{0}).
\end{equation}
\begin{figure*}[!htbp]
	\centering
\includegraphics[scale=0.6000,trim=1.5cm 0cm 3cm 0cm]{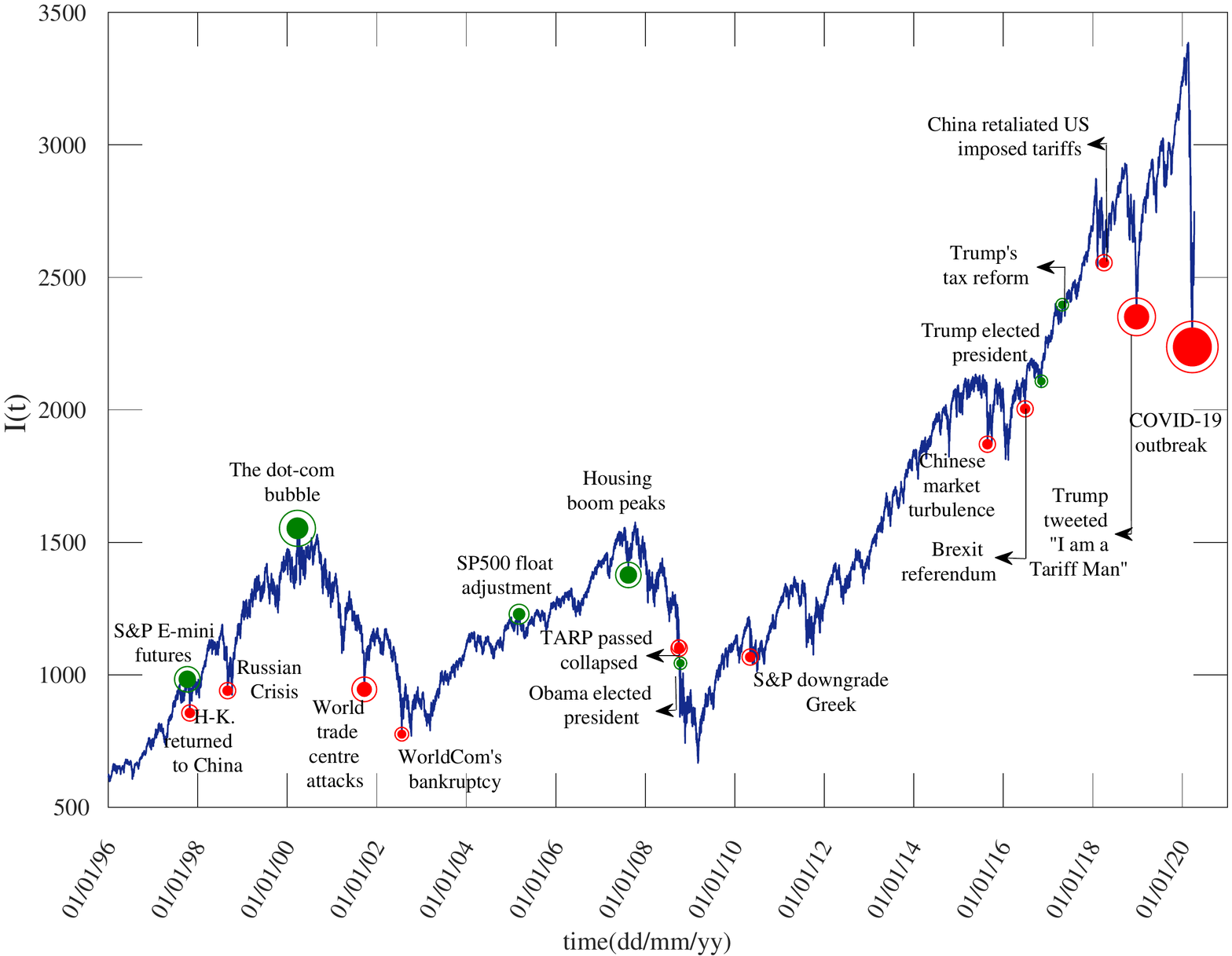}
\caption{The S\&P500 index $I(t)$ from 02/01/96 to 24/03/20 (24 years). In this data set the largest daily percentage loss of $-11.98\%$ was registered on 16/03/20. Some other key events are also shown for reference.}
\label{fig:Index}  
\end{figure*}
In earlier work we used a detrended fluctuation approach \cite{arias2019stationarity} to decompose the price return $X(t)$ into a deterministic component $\overline{X}(t)$ and a stationary {\it fluctuating component} $x(t)$  
\begin{equation}
X(t)=\overline{X}(t)+x(t).
\label{eq:detrended}
\end{equation}
The trend  $\overline{X}(t)$ was obtained by averaging the index over a moving time window. The size of the window was one year, that was optimized to guarantee that the fluctuations around the trend exhibit stationary behavior. \textcolor{black}{The governing  equation of this stationary behaviour is \cite{alonso2019q}:}
\textcolor{black}{
	\begin{equation}\label{ufo7}
	t^{1-\xi}\frac{\partial p}{\partial t}=\xi D \dfrac{\partial^{2}p^{2-q}}{\partial x^{2}},
	\end{equation}}
By using a curve-fitting analysis on the S\&P500 data over the past 24 years (see Appendix), we show that the probability density function (PDF) of the detrended price is well described by the functional form:
\begin{equation}
p(x,t)=\dfrac{1}{(Dt)^{1/ \alpha}} g_{q}\left( \dfrac{x}{(Dt)^{1/ \alpha}}\right) ,
\label{eq:P-Gaussian}
\end{equation}
\textcolor{black}{where $\alpha=\frac{3-q}{\xi}$. Being $D$, $q$ and $\alpha$  time-dependent fitting exponents which analysis is displayed in Figure \ref{fig:q_values}}. The $g_{q}$ term is the $q$-Gaussian function defined as:
\begin{equation}
g_{q}(x)=\dfrac{1}{C_{q}} e_{q}(-x^{2})
\label{eq:q-Gaussian}
\end{equation}

\noindent and the $q$-exponential function is $e_{q}(x) = \left[1+(1-q)x\right]^{\frac{1}{1-q}} $ that reduces to the exponential function when $q\to 1$. The normalization constant $C_{q}$ for $1<q<3$ is given by:
\begin{align}
Cq =\sqrt{\dfrac{\pi}{q-1}} \dfrac{\Gamma((3-q)/(2(q-1)) ) }{\Gamma(1/(q-1)) }.
\label{eq:Cq}
\end{align}
The cumulative distribution function (CDF) of the PDF Eq.(\ref{eq:P-Gaussian}) of the detrended price return is;
\begin{equation}
F(x,t)= \int_{-\infty}^{x} p(x,t) \, dx. 
\label{eq:CDF}
\end{equation}
The complementary of the CDF is used here to quantify risk under extreme events \cite{haas2009financial}. This is defined as probability that $X-\overline{X}$ lies outside the interval $[-x,x]$ is 
\begin{equation}
P( X(t)- \overline{X}(t) >|x| )=1-\int_{-x}^{x} p(x,t) \, dx.
\label{eq:PDF}
\end{equation}
\noindent For  this paper we introduce the standarization of the $q$-error function
\begin{equation}
\erf_{q}(x)=2 \int_{0}^{x} g_{q}(y)dy
\end{equation}
\noindent The standard error function is a special case of the $q$-error function for $q=1$. Considering this definition,  Eq.(\ref{eq:PDF}) is written as:
\begin{equation}
P( X(t)- \overline{X}(t) >|x| )=1-\erf_{q}(x/(Dt)^{1/\alpha}).
\label{eq:Probability}
\end{equation}

We use Eq.(\ref{eq:Probability}) to develop a forecast of the market based on the historical fluctuations in the S\&P500.  Figure \ref{fig:Index} shows some key dates over the last 24 years of the S\&P500. The market dynamics consists of periods of a bull market (systematic increase of the index) and a bear market (systematic decrease). The crashes in this plot are discontinuous. They occur in a very short period of time that we attribute as noise that {\color{black}it} is added to the deterministic component. For example, during the Global Financial Crisis (GFC) there is a long-term decline in the value of the S\&P500 that we can interpret as the deterministic trend, but near the point where Lehman Brothers collapses there is a discontinuous market crash. In fact, there are several very large daily price movements that are not always attributable to a specific event. These `crashes' (both up and down) are more similar to large stochastic movements, not a part of the deterministic component of the market dynamics.

\section{ Effect of COVID-19 to S\&P500} \label{Sec:Covid19}
\begin{figure}[!htbp]
	\centering
	\includegraphics[scale=0.4700,trim=1cm 0cm 3cm 5cm]{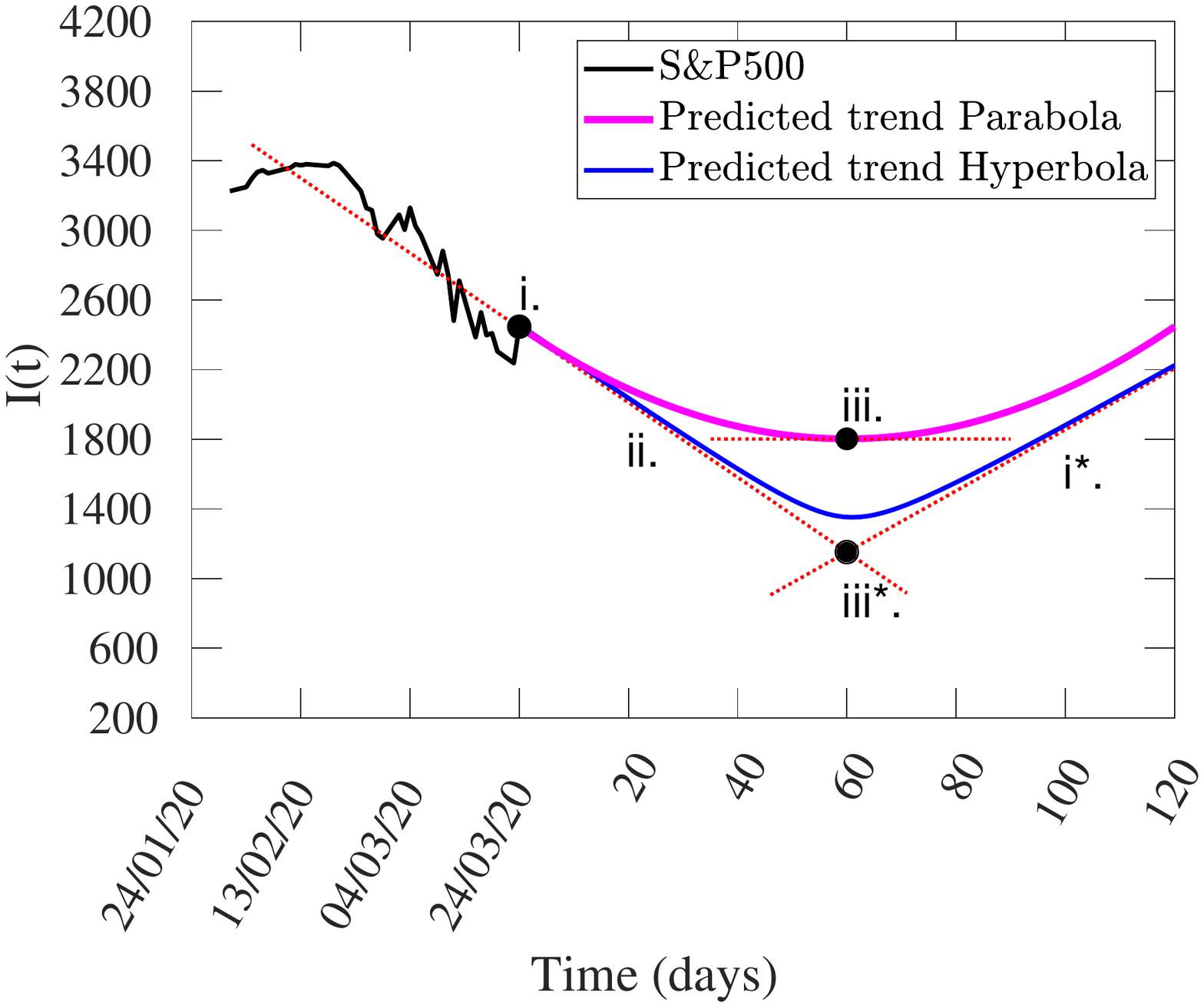}
	\caption{A downward trend followed by a market recovery
		is estimated by applying a parabolic fitting (magenta line) and hyperbola fitting (blue line). Three steps were applying to define each deterministic trend. The step ii is the same for both cases.} 
	\label{fig:Estimated_trend}      
\end{figure}

To establish the effect of COVID-19 on the market we assume that it has a deterministic, exogenous impact. This impact is assumed to be a response function that corresponds to a COVID-19 induced bear market followed by a smooth transition to a bull market. As a further refinement we assume that this transition is dictated by a key quantitative factor: the estimated date on which the mortality rate peaks. Our first estimation of this trend \textcolor{black}{(a private communication between authors on April 2, 2020)} was based on the time-frame from $31/01/2020$ to $24/03/2020$ in which a bear-market decline in $I(t)$ had already been observed. For the construction of the deterministic trend $\tilde{I}(t)$ the parabola and hyperbola functions were used. For the parabola three conditions were applied and are illustrated in Figure \ref{fig:Estimated_trend}: \textbf{(i)} The initial point of the predicted trend at $t_0$ is $\tilde{I}(t_0)$, \textbf{(ii)} the slope of $\tilde{I}(t)$ at $t_0$ is obtained from a linear fitting during the interval $31/01/2020$ to $24/03/2020$, \textbf{(iii)} the point where the recovery is predicted at $60$ days satisfies $\tilde{I}'(t_0 + 60 \, \textrm{days})=0$, i.e. 
\begin{equation}
\left. \frac{\tilde{I}(t)}{ d t}\right\vert_{t_0 + 60 \, \textrm{days}}  =  0.
\end{equation}
\\
For the hyperbola three similar conditions  were applied: The slope of the market recovery (bull market) \textbf{(i$^{*}$)} is assume to be 0.5 of the slope of the market's collapse similar to the previous crashes that have occurred over the past 24-years. The slope \textbf{(ii)} is the same as the parabola and point \textbf{(iii$^{*}$)} is the intersection of the slopes \textbf{(i$^{*}$)} and \textbf{(ii)}. For both methods the point \textbf{(iii)}  was based on the public information available by the University of Washington where the death rate was estimated to peak on 24/05/2020 (60 days from $t_0)$). To obtain this prediction of the trend we assumed that there is no time lag between the peak of mortality rate and the time where the market starts to recover.

\begin{figure}[!htbp]
	\centering
	\includegraphics[scale=0.37000,trim=0cm 0cm 4cm 1cm]{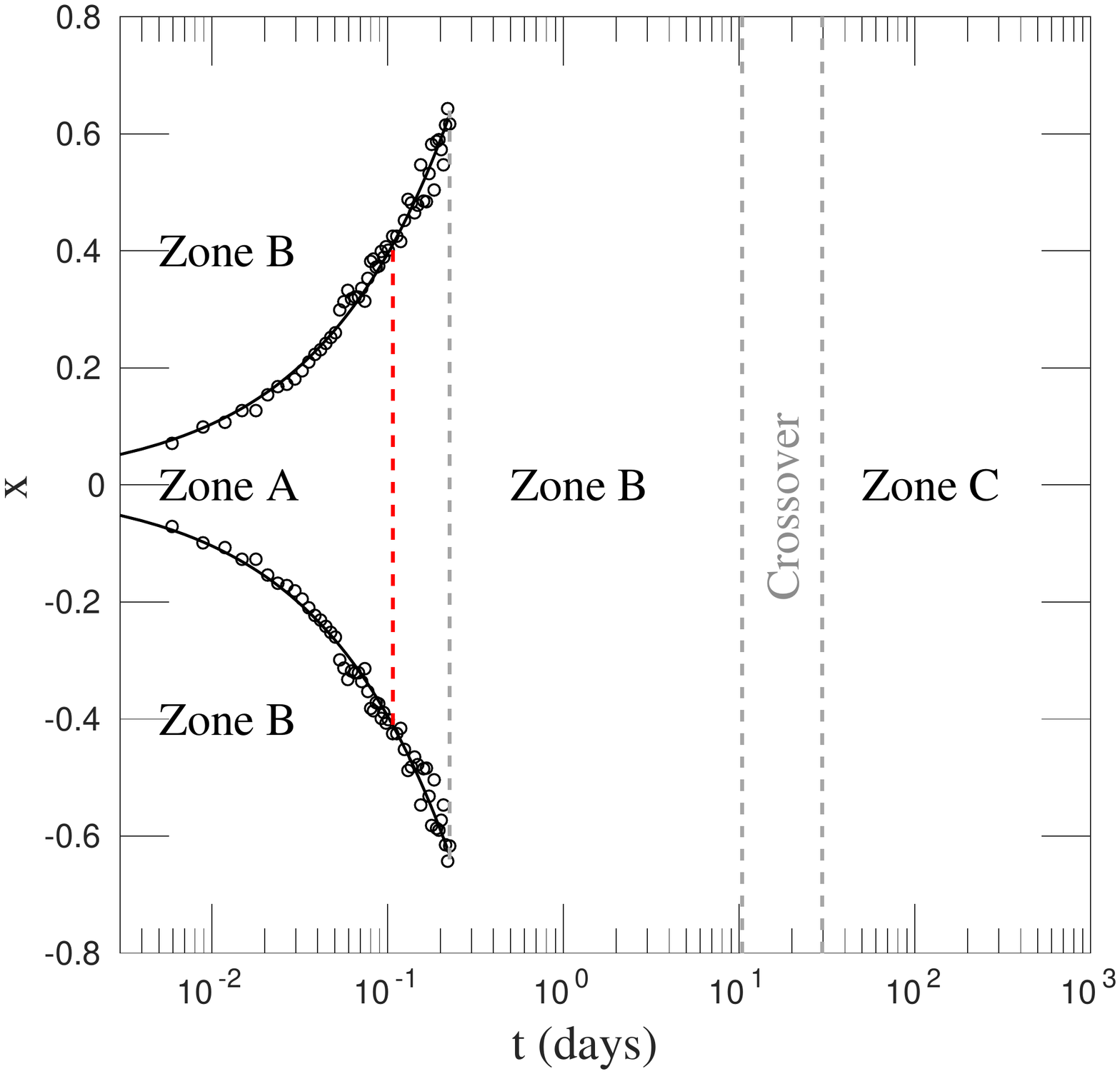}
	\caption{Three different zones were determined based on an abrupt slope change of $\alpha$ and $q$ values. The circles represent the end points of the strong super-diffusion regime (zone A) from $t_0$ to $t = 35$ minutes. The remaining area during the first $35$ minutes to $28$ days corresponds to a weak super-diffusion regime (zone B). A normal diffusion process is reach after thirty days. The gray dashed lines represent the transitions between each zones. }
	\label{fig:zone_values}       
\end{figure}
Up  to this point we have evaluated the systemic risk as the deterministic aspect of the market evolution, which is often neglected by analysts in the determination of the impact of exogenous events such as the COVID-19 pandemic. In what follows we evaluate the risk associated with the stochastic fluctuations of the index by `dressing' the deterministic risk with a $q$-Gaussian diffusion process. In a previous publication \cite{alonso2019q} we found that the $q$-Gaussian fluctuations of the S\&P500 index around the trend have distinctive super-diffusion spatio-temporal regimes (`space' refers to the size of the fluctuations). To allow long-term forecasting on the order of days, we have extended the analysis for longer times. Three well-defined regimes are observed in Fig. {\ref{fig:zone_values}},

\begin{itemize}
	\item Zone A: Strong superdiffusion process with short-time correlations
	\item Zone B: Weak superdiffusion process with weak time correlations
	\item Zone C: Normal diffusion process with no time correlations.
\end{itemize}

These zones are well-distinguished in time, where Zone A goes from 0 to 38 minutes, Zone B from 10 days to 28 days and after a crossover period Zone C starts at 30 days (Fig. {\ref{fig:zone_values}}). Note that the time series tends to a classical diffusion process for large times, in agreement with the classical Central Limit Theorem. See \cite{alonso2019q} for a complete analysis of these zones and their derivation. 
\par
\begin{figure}[!htbp]
	\centering
	\includegraphics[scale=0.4300,trim=1.5cm 0cm 0cm 1.0cm]{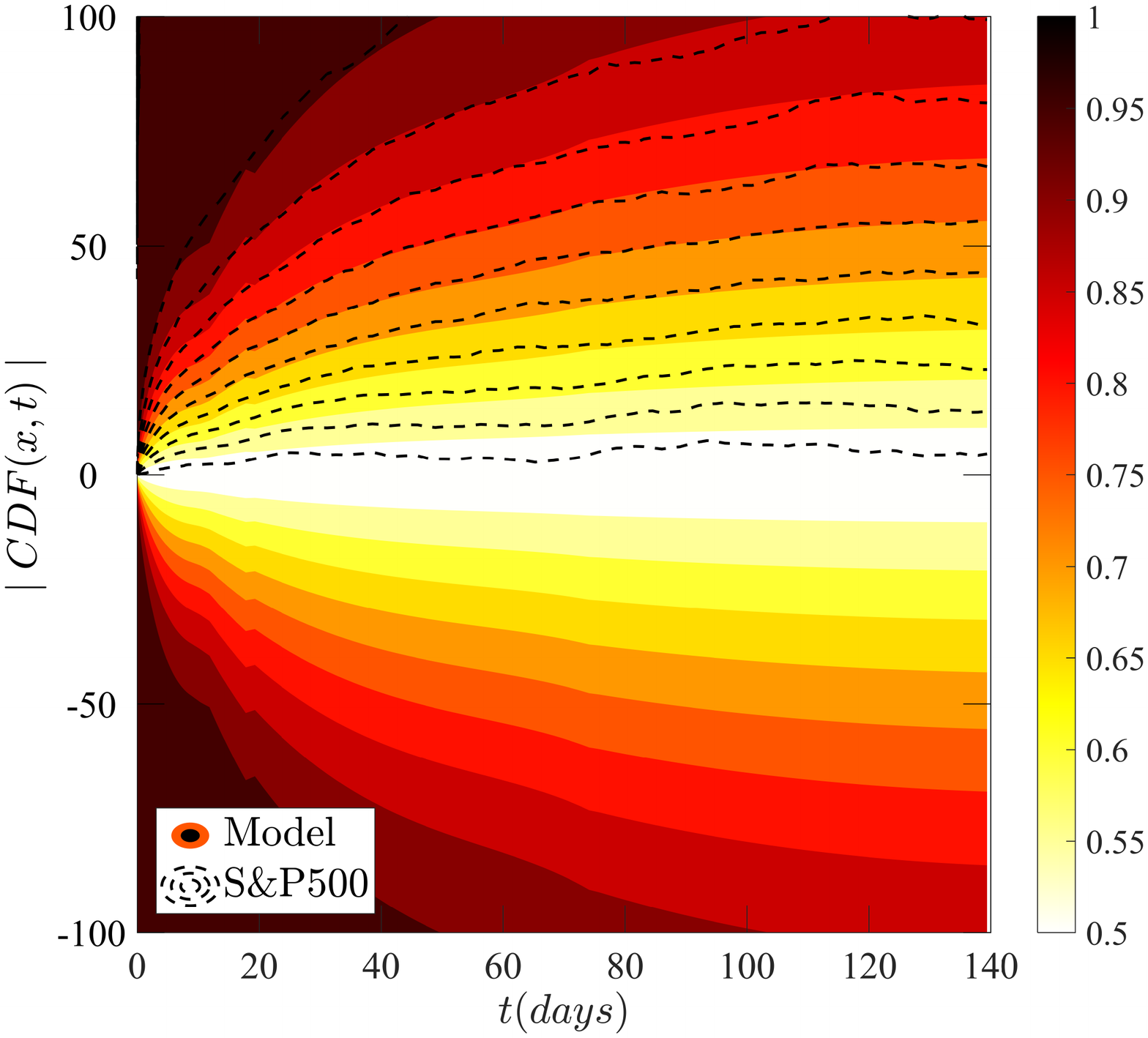}
	\caption{\textcolor{black}{Comparison of time evolutions of $CDFs$ between the price return of detrended stock market data and the analytical model which considers the time-dependent fitting parameter obtained from the second technique.}}
	\label{fig:Validation}       
\end{figure}

 
Two techniques were used to calculated the optimal $q$ for each $p(x,t)$, one focusing on the PDF and the other focusing on each CDF of \textcolor{black}{detrended price return}. For the first technique two methods were used. The least square method and  $q$-moments method. The least square method applies Eq.(\ref{eq:P-Gaussian}) as the fitting function. The $q$-moments method uses a system of two equations and two variables, $q$ and $\alpha$. The first equation is given by the \textit{``second moment"} or variance $\langle x^{2} \rangle$. The second equation refers to the \textit{``escort second moment"} or $q$-variance $\langle x^{2} \rangle_{q_{2}}$. In general, the $q$-moments are applied to PDFs with asymptotic decays because they provide finite values \cite{tsallis2009escort}.
The second technique is based on each CDF, where a least squares method is applied using Eq.(\ref{eq:CDF_sol}). \textcolor{black}{The quality of the fitting models was evaluated after reproducing the time evolution of $CDF(x,t)$ with the time-dependent fitting parameters previously obtained, then we compare them with the $CDF$ of detrended price return. The best results were obtained with the second technique, which displays a  smoothness and more accurate time evolution of the $CDF$ of detrended price (Figure \ref{fig:Validation}).}

We extend the analysis of this derivation in the supplementary materials section.

The corresponding $q$, $\alpha$ and $D$ values that fit each of these equations are displayed in Fig.{\ref{fig:q_values}}(a-c-d). The $q$ and $\beta$ values were calculated directly after fitting Eq.(\ref{eq:CDF_sol}). The $\alpha$ values are calculated as the power law of the $\beta$ value, by averaging the power law over a moving time window smaller than the transition zone. The $D$ value is calculated by replacing $\alpha$ in Eq.(\ref{eq:beta}). The results for $q$, $\alpha$ and $D$  obtained by applying the least squares method of the CDFs are consistent with the ones calculated focused on the PDFs. The convergence of $q\rightarrow1$ and $\alpha\rightarrow2$ shows that the PDF of $x$ is Gaussian when $t\rightarrow\infty$.

\begin{figure}[!htbp]
	\centering
	\includegraphics[scale=0.4300,trim=0.5cm 0cm 0cm 1.0cm]{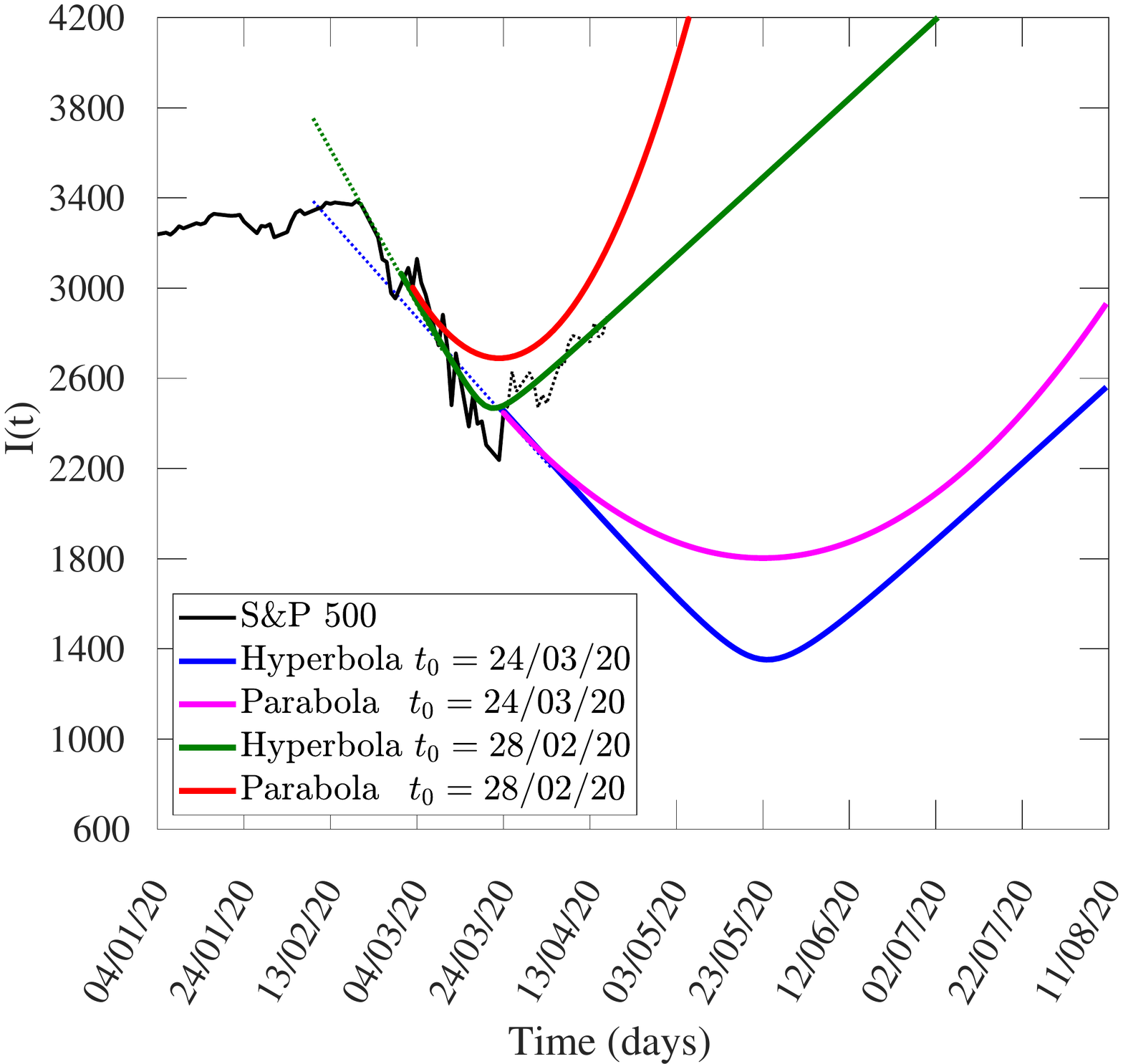}
	\caption{The decline and recovery on four different trends due to COVID-19 are observed. The first forecast method considers $t_0=24/03/2020$ and the second one $t_0=28/02/2020$. For both cases a parabola and hyperbola fitting was used, considering the method previously illustrated in Fig. \ref{fig:Estimated_trend}. The hyperbola method with $t_0=28/02/2020$ displays the most accurate result at the moment.}
	\label{fig:Trend_and_fluctuations}       
\end{figure}

Then, we construct a forecast of S\&P500 using two predicted trends, the parabola and the hyperbola shown in Fig.\ref{fig:Trend_and_fluctuations}. We note that these trends are approximations based upon the prices for the following trading days. For that reason both trends were re-calculated using $t_0=28/02/2020$; the date on which the first death occurred in USA; and $\tilde{I}'(60)=0$; two months after the first case was confirmed. These new trends display a better performance, notably the hyperbola which fits better than the others.

The uncertainty was modeled with the analytical form of the detrended price by replacing the  $\alpha(t)$, $q(t)$, and $D(t)$ calculated in the Appendix into Eq.(\ref{eq:P-Gaussian}) with their numerical estimates. Once the stochastic term is simulated it is added into the deterministic trend to see the full stochastic path of $I(t)$.
\par 

\begin{figure}[!htbp]
	\centering
	\includegraphics[scale=0.4300,trim=1.5cm 0cm 0cm 1cm]{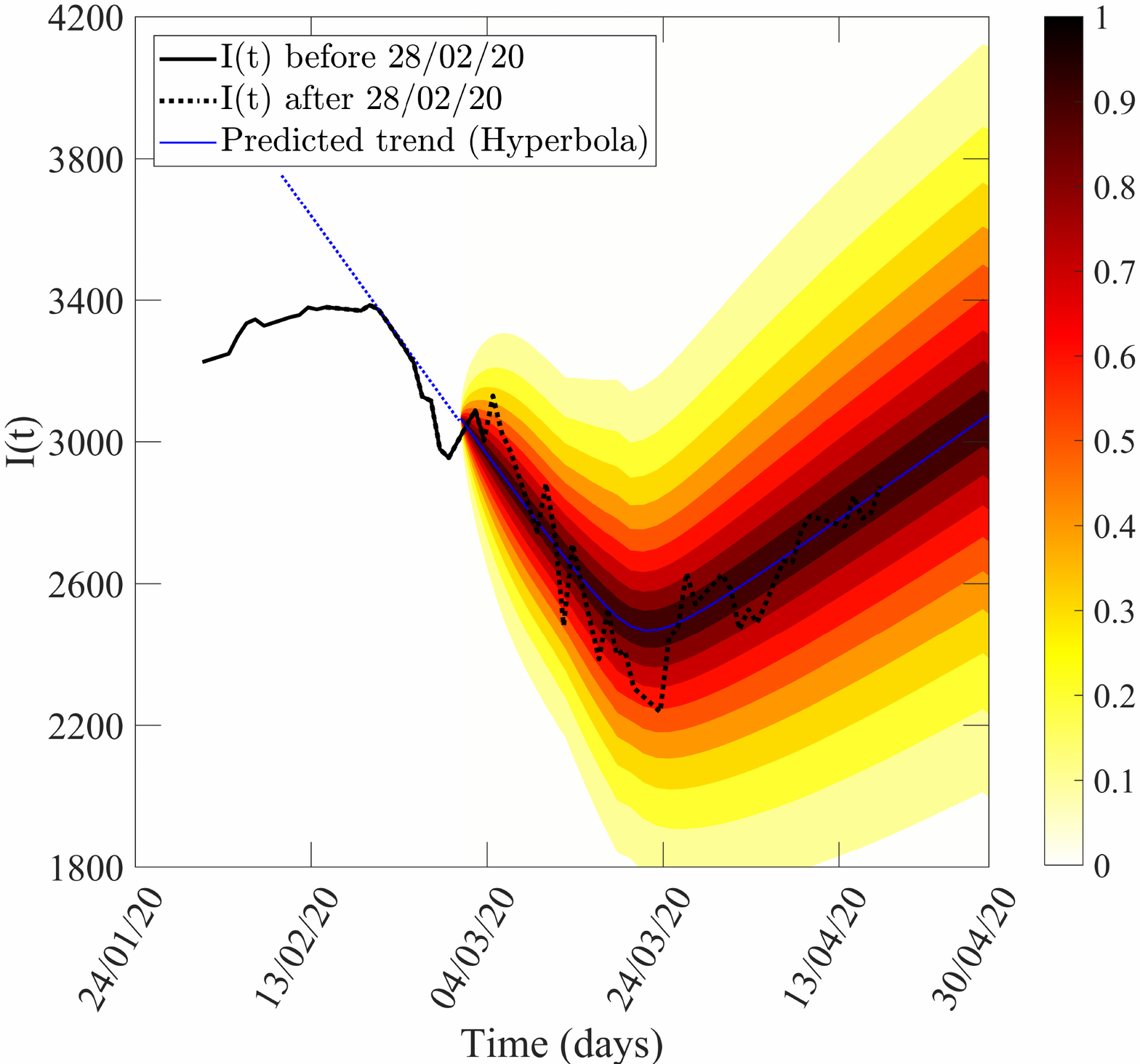}
	\caption{The decline and recovery on $I(t)$ due to COVID-19 is observed from $28/02/20$ to the following $2$ months. A downward and upward market predicted trend is  calculated by applying a hyperbolic fitting with no time lag. The uncertainty is represented by a scale from $0$ to $1$, where represent the $P( X- \overline{X} >|x| )$ of each contour line. The economy will start its recovery after this epidemic peak of deaths}
	\label{fig:Forecast}       
\end{figure}
Figure \ref{fig:Forecast} shows the forecast result identifying the range of possibilities during the market's decline and subsequent recovery. The predicted trend was plotted as a hyperbola with $t_0=28/02/2020$ (Figure \ref{fig:Trend_and_fluctuations}). The uncertainty is presented by a shaded contour plot with a scale from 0 to 1. These values represent the probability of a possible variation of $P( X- \overline{X} >|x| )$ which can be visualized as a ``cone of uncertainty''. This probabilistic path of the S\&P500 over the following 60 days (2 months) shows the evolution of the PDF of $I(t)$, which diffuses as $t$ increases. {\color{black}Also we see that the real data for index after $t_0$ (the dashed line) lies within a tiny interval in the vicinity of the predicted trend, mainly in the most probable (dark) regions, verifying that our model is working and is reliable for forecasting with $ 85\% $ of accuracy.}\par

We construct a forecasting based on essential public information provided by the University of Washington on their web page\cite{WHS}. A good forecasting is always an iterative process that can be daily updated considering new factors and changes. 

\section{Conclusions}\label{Sec:Discussion}

In summary, we have presented a model of the stochastic and systematic risk in the stock market and applied it to forecast the stock market response to COVID-19. We have assumed that the stochastic risk is a $q$-Gaussian diffusion process. The systemic risk is the deterministic aspect of the market evolution that is often neglected by market analysts  but here we have assumed that COVID-19 has a deterministic, exogenous impact on the market. We have dressed this nonlinear skeleton with a $q$-Gaussian diffusion process that we developed in previous work. The response function we have used for forecasting is a simple behavioural model based on the assumption that markets recover once prices drop to sufficiently low enough levels, in future work it can be improved by developing better tools to evaluate systematic risk using behavioural~\cite{bikhchandani2000herd} or macroeconomic~\cite{gertler2016wholesale} models. In addition, our stochastic method is still heuristic and it needs to be developed by formulating the governing equations that allows transition from strong- to weak- superdiffusion and converging to a normal diffusion process for large times, as dictated by the classical Central Limit Theorem.
\textcolor{black}{However, this method opens up a potential opportunity for risk control in other areas such as climatology, seismology (earthquakes) and communication networks, where large multivariate data sets provide a window on the dynamics of the underlying system.} 

\section{ACKNOWLEDGMENTS}
We acknowledge the Australian Research Council grant DP170102927. K.A.C. thanks The Sydney Informatics Hub at The University of Sydney for providing access to HPC-Artemis for financial data processing. We thanks Sornette, Constantino Tsallis and Christian Beck for inspiring discussions.

\bibliographystyle{rsc}
\bibliography{GoverningEquation}

\appendix
\section{Supplementary materials}

\counterwithin{figure}{section}

The probability density function (PDF) of the detrended price is well described by the functional form:
\begin{equation}
p(x,t)=\sqrt{\beta}\, g_{q}\left( { \sqrt{\beta}}x\right),
\label{eq:P-Gaussian_beta}
\end{equation}
where, $\beta=(D t)^{-2/\alpha}$, allows to recovery Eq.(\ref{eq:P-Gaussian}).

Then, a system of two equations based on variance $\langle x^{2} \rangle$  and $q$-variance $\langle x^{2} \rangle_{q_{2}}$ were used to obtain the $q$ and $\beta$ values. \par

The variance $\langle x^{2} \rangle$ has a finite value for $q<5/3$;
\begin{equation}\label{eq:System}
\langle x^{2} \rangle =\dfrac{1}{\beta(5-3q)}.
\end{equation}

The normalized $q$-variance $\langle x^{2} \rangle_q$ is defined as \cite{tsallis2009escort}
\begin{equation}\label{eq:q-Moments}
\langle x^{2} \rangle_{q}=\dfrac{\int_{-\infty}^{\infty} x^{2} p(x,t)^{q} dx}{\int_{-\infty}^{\infty} p(x,t)^{q} dx},
\end{equation}
In general, the q-moments are calculated on PDFs with heavy tails because they provide finite values.
The q-moments and the standard moments are equal for $q=1$.\par
{\color{black}To obtain the analytical solution of the integral in the numerator and denominator of Eq.(~\ref{eq:q-Moments}), we notice that 
	\begin{equation}\label{eq:New_Integrand}
	K(\beta,q)\equiv\int_{-\infty}^{\infty}[1-(1-q) \beta x ^{2}]^{\dfrac{1}{1-q}}dx=\dfrac{C_{q}}{\sqrt{\beta}},
	\nonumber
	\end{equation}
	and also the second moment of $x$ is proportional to $\partial \beta K(\beta,q)$, so the analytical solution of the numerator is:}

\begin{equation}\label{eq:New_Integrand_Sol}
\begin{split}
\dfrac{\partial}{\partial \beta} K(\beta,q)=&-\int_{-\infty}^{\infty} x^{2}[1-(1-q)\beta x^{2}]^{\dfrac{q}{1-q}}dx\\=&-\dfrac{1}{2} \beta^{-3/2} C_{q}
\end{split}
\nonumber
\end{equation}
Then, the analytical solution of the denominator is:
\begin{equation}
{\int_{-\infty}^{\infty} p(x,t)^{q} dx}= \dfrac{3-q}{2}
\nonumber
\end{equation}
\begin{figure*}[!htbp]
	\centering
	\includegraphics[scale=0.40,trim=1.5cm 0cm 10cm 10.5cm]{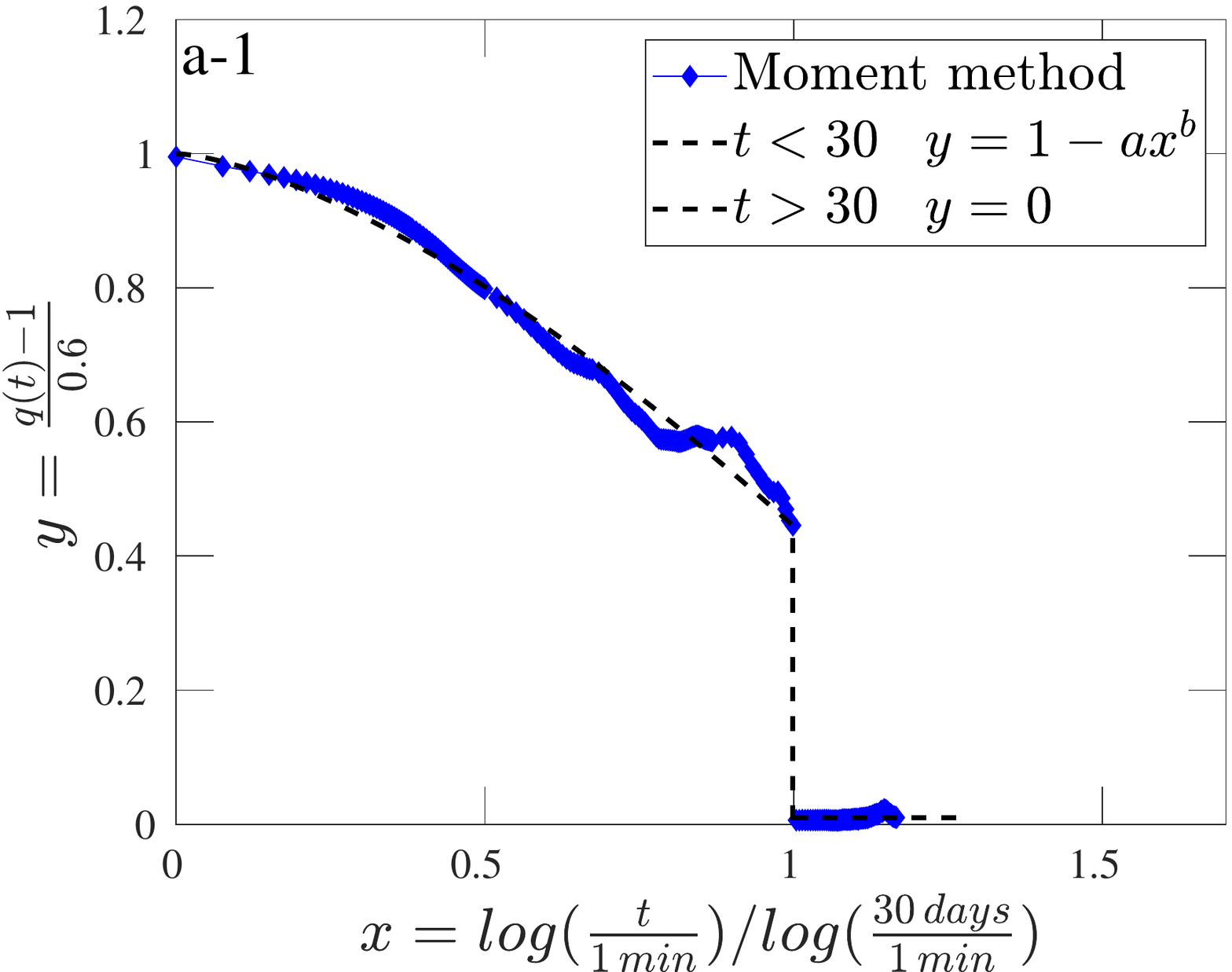}
	\includegraphics[scale=0.40,trim=1.5cm 0cm 10cm 10.5cm]{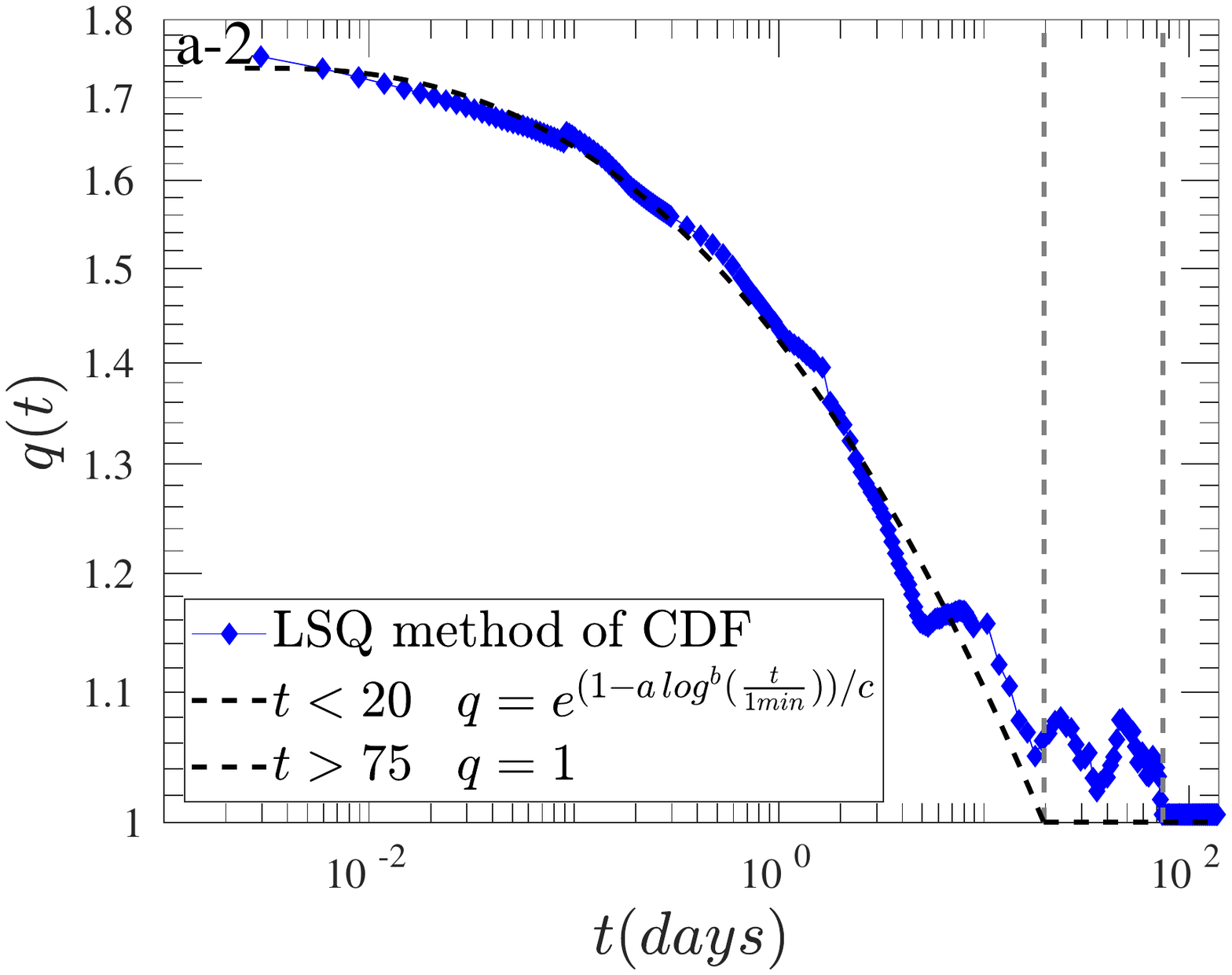}
	\includegraphics[scale=0.40,trim=1.2cm .3cm 10cm 6.0cm]{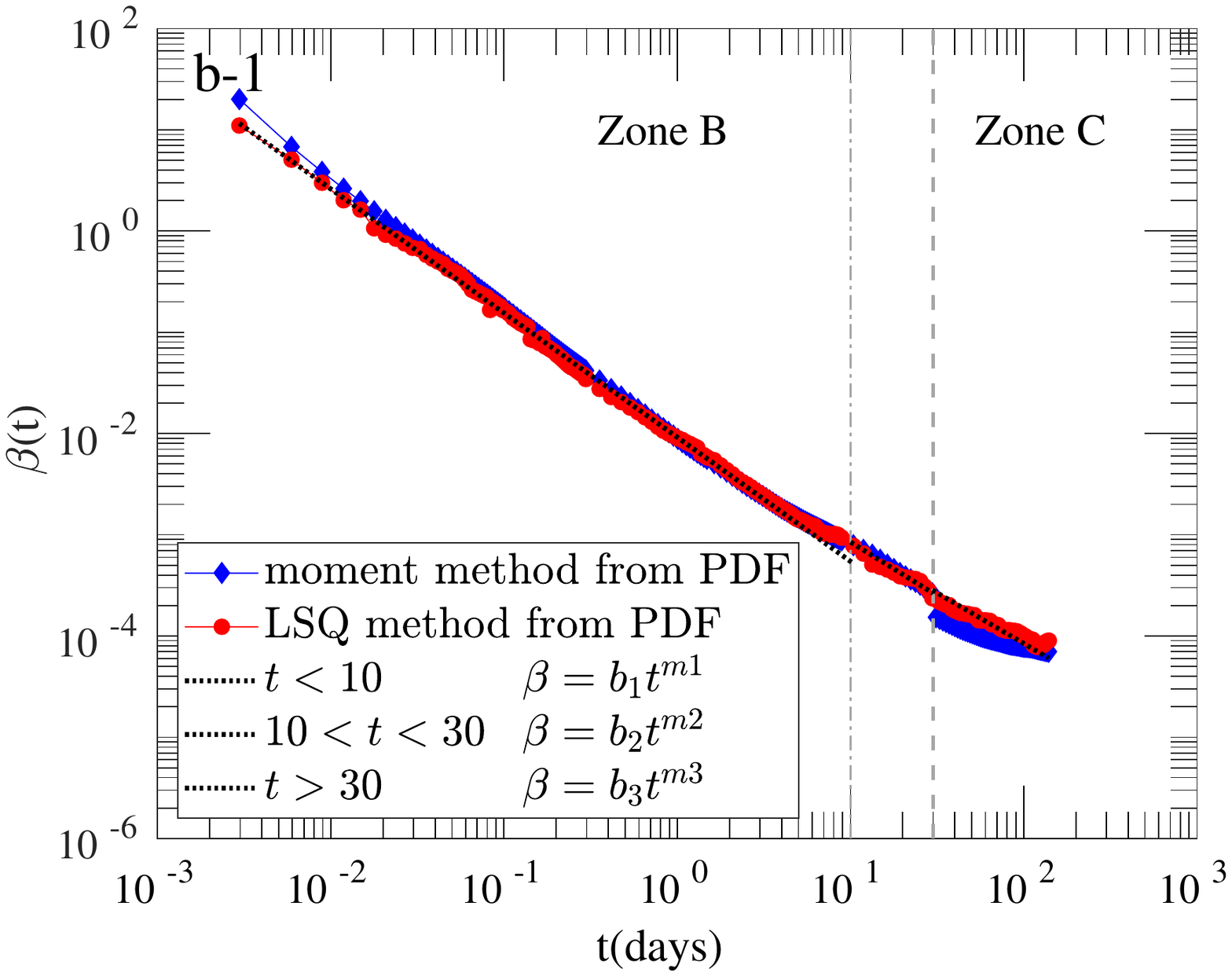}
	\includegraphics[scale=0.40,trim=1.2cm .3cm 10cm 6.0cm]{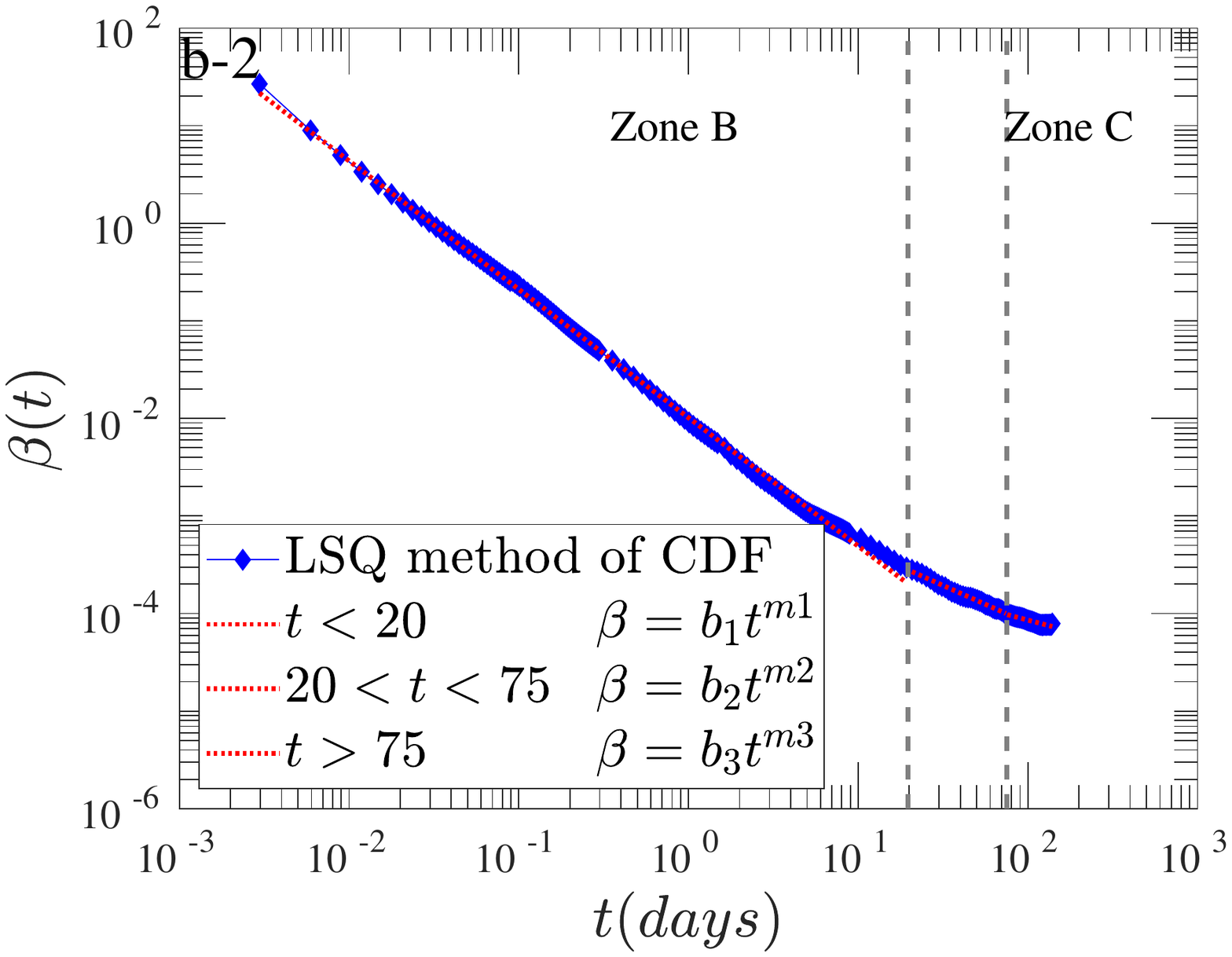}
	\includegraphics[scale=0.40,trim=1.2cm 0cm 10cm 6.5cm]{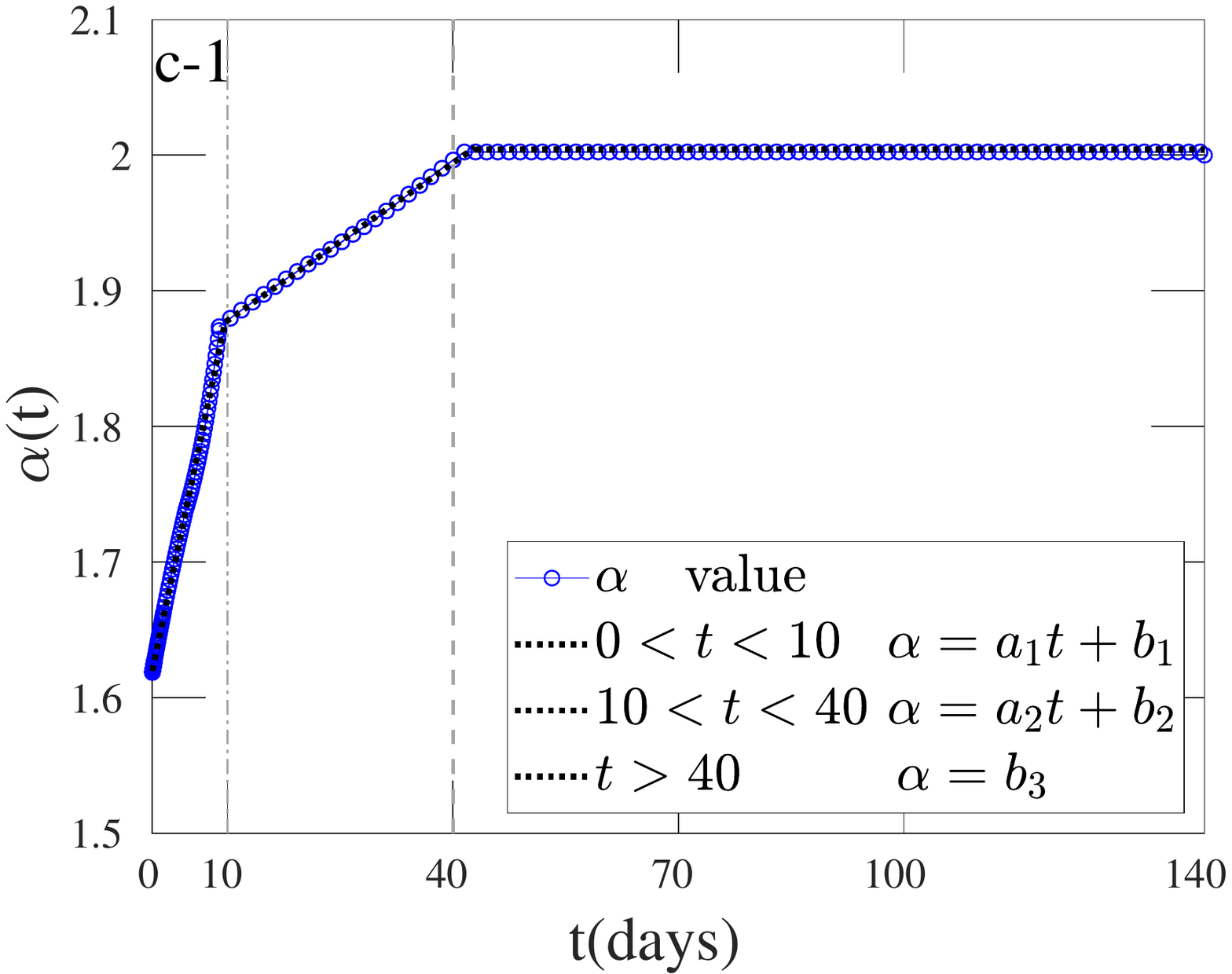}	\includegraphics[scale=0.40,trim=1.2cm 0cm 10cm 6.5cm]{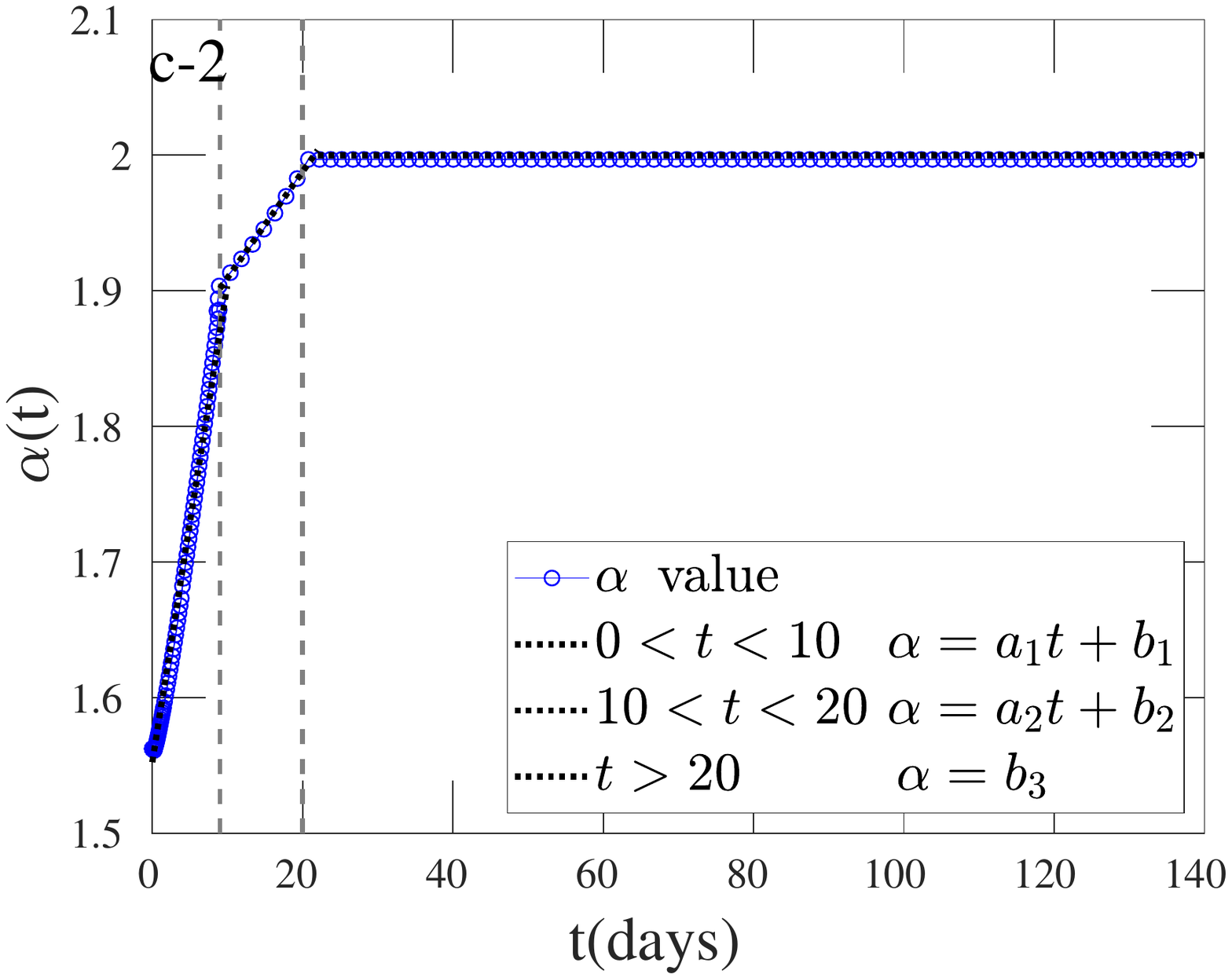}
	\includegraphics[scale=0.40,trim=1.2cm 0cm 10cm 6.5cm]{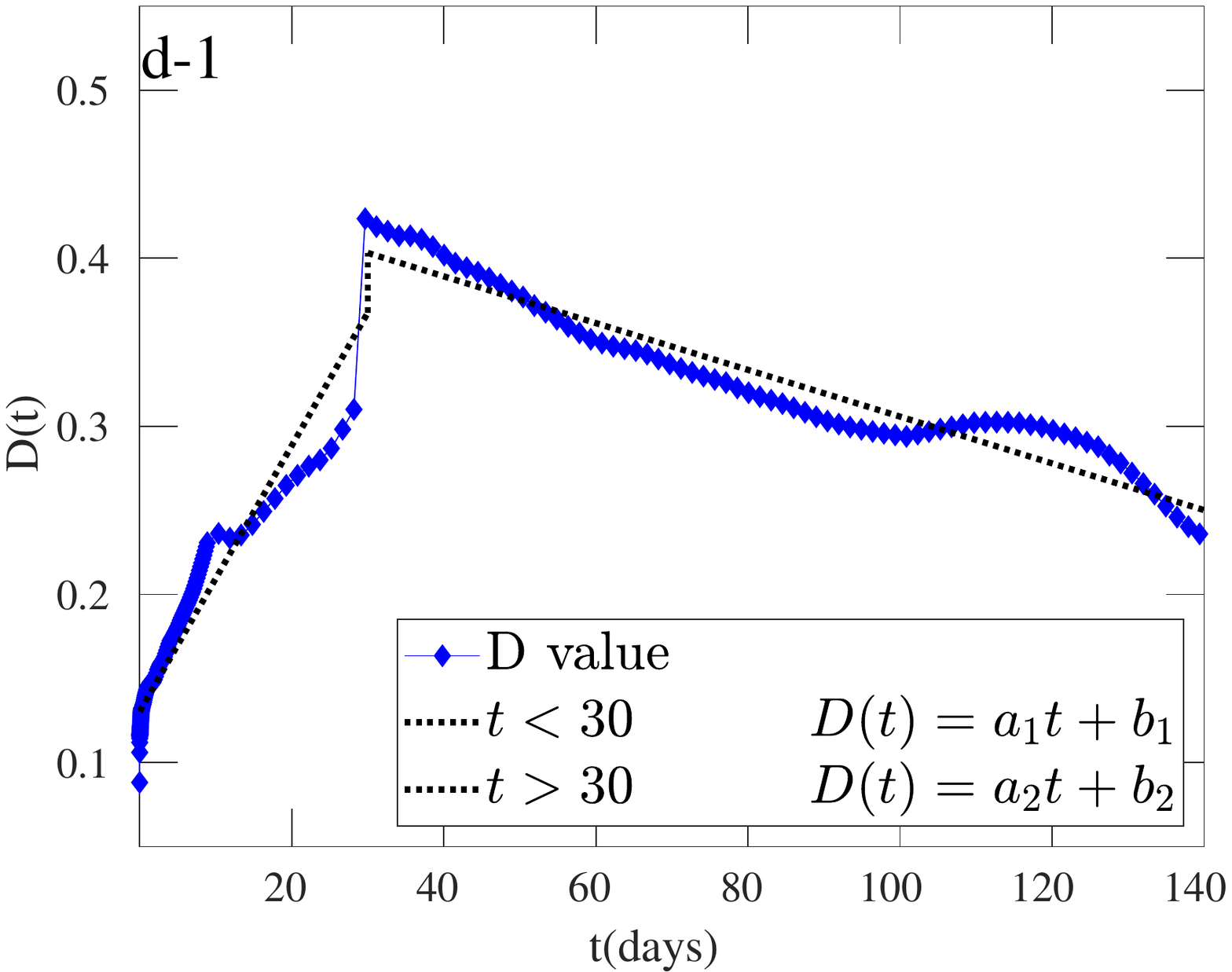}
	\includegraphics[scale=0.40,trim=1.2cm 0cm 10cm 6.5cm]{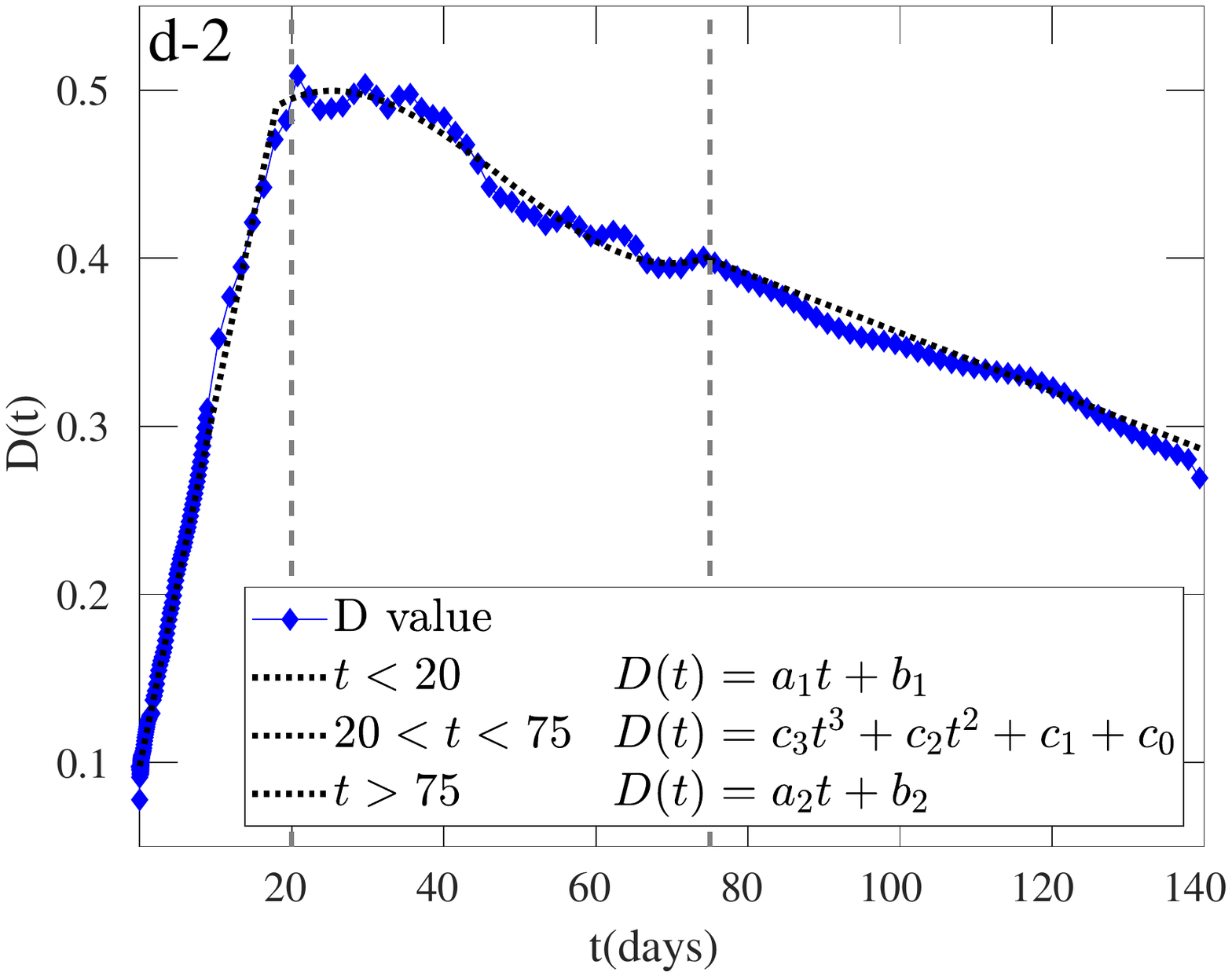}
	
	\caption{Calculation of $q$ and $\beta$ by applying a Equation system  [Eqs(\ref{eq:System}) and (\ref{eq:System2})] and fittings of $F(x,t)$. (a-1, a-2) Estimation of $q(t)$ value, for (a-1) a normalization of units was made. (b-1,b-2) The value of $\beta\sim t^{-2/\alpha}$. (c-1,c-2) The exponent of the power law relation of $\beta$ vs $t$ per time window is calculated as $\alpha(t)$. (d-1,d-2) The $D$ value is calculated by replacing $\alpha$ in Eq.(\ref{eq:beta}). The convergence to a Normal distribution function is observed when $t\rightarrow\infty$ with $q=1$ and $\alpha=2$. (coefficients are in Table \ref{Tab:Fitting}) }
	\label{fig:q_values}       
\end{figure*}

\begin{table*}[!htbp]
	\caption{Coefficients values of fittings}
	\begin{center}
		\begin{tabular}{|c|c|c|} 
			\hline
			& Fitting $p(x,t)$  & Fitting $F(x,t)$ \\ 
			\hline
			& $a=0.556\pm0.012$ & $a=0.005\pm 0.001$\\ 
			$q$ & $b=1.485\pm0.041$ & $b=2.473\pm0.158$\\ 
			&          & $c=1.811\pm0.031$\\ 
			\hline
			& $b_{1}=0.009\pm3.09(10)^{-4} \quad m_{1}=-1.23 \pm 0.01$ & $b_{1}=0.010\pm 3.5 (10)^{-5}  \quad m_{1}=-1.32\pm 0.01$\\ 
			$\beta$ & $b_{2}=0.008\pm3.5(10)^{-5} \quad m_{2}=-1.10\pm 0.01$ & $b_{2}=0.003\pm7.5(10)^{-4} \quad m_{2}=-1.23\pm0.05$\\ 
			& $b_{3}=0.010\pm2.66(10)^{-4} \quad m_{3}=-1.06\pm0.02$ & $b_{3}=0.007\pm8.77(10)^{-4} \quad m_{3}=-0.97\pm0.01$\\ 
			\hline
			& $a_{1}=0.027 \pm{0.001} \quad b_{1}=1.620\pm0.003$ & $a_{1}=0.035\pm0.001 \quad b_{1}=1.553\pm0.002$ \\ 
			$\alpha$ & $a_{2}=0.004\pm{0.002} \quad b_{2}=1.839\pm{0.009}$  & $a_{2}=0.008\pm0.001 \quad b_{2}=1.831\pm0.007$\\ 
			&  $ b_{3}=2.00\pm0.000$ & $ b_{3}=2.00\pm0.000$\\ 
			\hline
			& $a_{1}=0.008\pm0.015 \quad b_{1}=0.130\pm0.002$ & $a_{1}=0.022\pm0.010 \quad b_{1}=0.098\pm0.001$ \\ 
			$D$ &$a_{2}=-0.001\pm7.14(10)^{-5} \quad b_{2}=0.445\pm0.004$  & $c_{3}=2.308(10)^{-6}\pm0.765(10)^{-7} $  \\ 
			&  &$c_{2}=-3.30(10)^{-4}\pm1.00(10)^{-5} $ \\ 
			&  &$c_{1}=0.012\pm0.005 \quad c_{0}=0.363\pm0.072$ \\ 
			&  & $a_{2}=-0.002\pm6.85(10)^{-5} \quad b_{2}=0.525\pm0.007$ \\ 
			\hline
		\end{tabular}
		\label{Tab:Fitting}
	\end{center}
\end{table*}
{\color{black}Inserting this identity into the Eq.(\ref{eq:q-Moments}), we obtain}:
\begin{equation}\label{eq:System2}
\langle x^{2} \rangle_{q_{2}}=\dfrac{1}{(3-q) \beta} .
\end{equation}
The equation system is defined by Eqs(\ref{eq:System}) and (\ref{eq:System2}) and is used to calculate $\beta$ and $q$ as a function of time. The $\alpha$ and $D$ values are calculated using:
\begin{equation}\label{eq:beta}
\beta=(D t)^{-2/\alpha}
\end{equation}
We divide the data of S\&P500 in overlapping time windows not longer than the transition zones. In each window the exponent of the power law relation is calculated and then the Eq.(\ref{eq:beta}) is used to calculate {\color{black}$D$ and $\alpha$}. The results for $q$,$\alpha$ and $\beta$ are shown in Subfigures a-1,b-1,c-1 and d-1.


In Subfigure a-1 an abrupt transition in $q$ is noticed when $t=30$ days. This abrupt change leads to a discontinuity in the fitted function $q(t)$. \\
{\color{black}We have  used an alternative method to extract the parameters, for which this discontinuity is absent.} This method is based on the fitting of the $F(x,t)$ of real price return $x$. The analytical expression of $F(x,t)$ is obtained in terms of the $\erf_{q}(x,t)$. First, Eq.(\ref{eq:CDF}) is replaced in Eq.(\ref{eq:PDF}),
\begin{equation}
P( X(t)- \overline{X}(t) >|x| )=2-2F(x,t).
\label{eq:CDF_in_PDF}
\end{equation}

{\color{black}Then, Eq.(\ref{eq:CDF_in_PDF}) and Eq.(\ref{eq:Probability}) are compared, resulting to ($s=x \sqrt{\beta}$)};
\begin{equation}
F(x,t)=0.5+\dfrac{\erf_{q}(s)}{2}
\label{eq:CDF_sol}
\end{equation}

Where, 
\begin{equation}
\erf_{q}(s)=\dfrac{2 \sqrt{\beta}}{C_{q} }  x \,\,_{2}F_{1} \left[ \dfrac{1}{2}, \dfrac{1}{q-1},\dfrac{3}{2},(1-q) s^{2}\right]. 
\label{eq:Erf}
\end{equation}
The Eq.(\ref{eq:CDF_sol}) is used to calculate $q$ and $\beta$ value as a function of time. The $\alpha$ values are calculated again with Eq.(\ref{eq:beta}) by overlapping time windows, and then, $D$ is obtained by replacing $\alpha$ in \ref{eq:beta}.\\ These results are shown in the second column of Figure \ref{fig:q_values}.
The Subfigure (a-2) does not display an abrupt transition of $q$ value. For $\beta$, $\alpha$ and $D$ the transitions are smooth in each zone.

\printcredits


\end{document}